\documentclass[runningheads]{llncs}
\usepackage[T1]{fontenc}
\usepackage{booktabs} 
\usepackage{graphicx}
\usepackage{multirow}
\usepackage{subcaption}
\usepackage{tablefootnote}
\usepackage{tablefootnote}

\usepackage{color}
\usepackage{amssymb}
\usepackage[most]{tcolorbox}
\begin{document}

\newtcolorbox{boxA}{
    fontupper = \scriptsize\bf,
    boxrule = 1pt,
    colframe = black 
}
\title{Security in a Workflow: Exploring Role-Based Agentic Architectures for Vulnerability Handling}

\titlerunning{Security in a Workflow}
%
\author{Srijita Basu\inst{1}\orcidID{0000-0002-6835-947X} \and
Miroslaw Staron\inst{1}\orcidID{0000-0002-9052-0864}}
%
\authorrunning{Basu \& Staron}
%
\institute{Chalmers University of Technology
and University of Gothenburg,
417 56 Göteborg, Sweden
\email{\{srijita.basu,miroslaw.staron\}@gu.se}}
\maketitle              
%


\begin{abstract}
Secure software engineering in practice is a multi-stage workflow involving vulnerability analysis, remediation, and fix verification. However, current LLM-based software security approaches often focus on isolated tasks such as detection or patch generation, with limited attention to agentic architectures reflecting industrial workflow. This creates a gap between existing LLM-based vulnerability-handling methods and real-world practices. In this paper, we study a role-based agentic workflow for vulnerability analysis and mitigation consisting of Planner, Analyzer, Fixer, and Verifier roles. To explore the effect of static-analysis tool, the analyzer agent was integrated with the CodeQL in one of the workflows. The models used include \textit{nemotron-cascade-2:30b, qwen3-coder-next, and gpt-oss:120b}. Our evaluation uses 25 real-world C/C++ vulnerabilities. The study reports 44\%  vulnerability detection accuracy comparable to GPT 5.5 and 19\% fix accuracy. We also list implications from this study in context of software security practitioners.

\end{abstract}

%
%
%
\section{Introduction}
Secure software engineering has become an essential part of modern software development, as security vulnerabilities directly affect quality, reliability, and trust. In industrial practice, vulnerability handling is typically not a single-step task but a structured workflow involving vulnerability analysis, remediation, and fix verification. With the benefits of large language models (LLMs), LLM-assisted cybersecurity capability is advancing rapidly, e.g., XBOW topped the HackerOne US leaderboard \footnote{https://xbow.com/blog/top-1-how-xbow-did-it}, Google Big Sleep finds real-world vulnerabilities showing that LLMs are moving from synthetic evaluation towards operational bug discovery \footnote{https://therecord.media/google-big-sleep-ai-tool-found-bug},DARPA AI Cyber Challenge (AIxCC) at DEF CON 33 (security conference) exhibited high efficiency in vulnerability detection and patching \footnote{https://aicyberchallenge.com/}. 


However, current research remains fragmented. Existing studies often focus on isolated tasks such as vulnerability detection \cite{ref_article1}, repair benchmarking \cite{ref_article2}, or patch generation \cite{ref_article3}, while comparatively limited attention has been paid to workflow-oriented agentic architectures \cite{ref_article4} that reflect how security activities are organized in real software engineering processes. 
Prior work in LLM-based software engineering suggests that agent collaboration quality depends on factors such as role alignment, interaction patterns, and consistency between generated code and accompanying reasoning~\cite{basu2026multiagent,ieee_literate_programming_llms}. However, these insights have largely been studied in general software engineering settings, with much less attention to security aspects. This motivates examining whether similar structured agentic designs are useful in secure software engineering.

In addition, there is still limited work exploring tool-augmented agentic workflows and Model Context Protocol (MCP)-style skill integration for low-level systems languages such as C and C++, which remain widely used in safety-critical systems. Compared with high-level languages such as Python or Java, vulnerabilities in C/C++ often depend on low-level memory semantics, pointer manipulation, buffer management, and execution behavior, making both diagnosis and remediation more challenging for LLMs \cite{ref_article5}. This creates a gap between current LLM-based vulnerability handling methods and the needs of practical secure software engineering, particularly in settings where contextual reasoning, explainability, transparent outcomes, and reliable tool support are essential.

In this paper, we investigate a role-based agentic workflow for vulnerability analysis and mitigation as an initial security-focused software process improvement mechanism for secure software engineering. In this context, process improvement means making security activities more structured, repeatable, transparent, and reviewable by decomposing vulnerability handling into explicit stages such as investigation planning, vulnerability analysis, patch generation, and fix verification. We instantiate the workflow in CrewAI and compare two variants: i) Planner, Analyzer, Fixer \& Verifier and ii) Planner, Analyzer with MCP CodeQL integration, Fixer \& Verifier. We conducted a focused exploratory study of whether workflow-oriented role separation and the addition of static-analysis capability at the analysis stage can improve the quality of vulnerability detections, generated patches, and verification outcomes. The evaluation is performed on 25 curated real-world C/C++ Common Vulnerabilities and Exposures (CVEs) spanning five Common Weakness Enumeration (CWE) categories (CWE-787, CWE-476, CWE-190, CWE-125, CWE-191). The selection was based on frequently encountered CWEs as elaborated later, and for each CWE, the most recent CVE instances were selected for which we could extract enough information required for formulating the ground-truth comprising of the vulnerability details from Natioanl Vulnerability Database (NVD) and patched version of the code provided in the repository under study. This was necessary for later evaluation.

This study is motivated by two research questions: 
1) \emph{RQ1}: How effective is the proposed role-based agentic workflow as an initial exploratory mechanism for studying process-oriented vulnerability analysis, patch generation, and fix verification? 2) \emph{RQ2}: How does integrating MCP tool based skills into the Analyzer stage affect the quality of downstream workflow outcomes, including detections, generated patches, and verification results?

We position this work as a methodology-oriented exploratory study rather than a large-scale benchmark. The smaller curated dataset was chosen deliberately to support detailed stage-wise analysis of the proposed workflow, including how planner guidance, role separation, and CodeQL-supported diagnosis affect vulnerability analysis, patch generation, and verification. The main contributions of this paper are as follows: \emph{1) Propose and evaluate an end-to-end role-based agentic architecture for vulnerability detection, patch generation, and fix verification, 2) Design a static Planner module for guiding vulnerability investigation in industry-relevant C projects. The results largely vary with and without the planner guidance, \& 3) Provide an initial curated reference set of 25 real-world C/C++ vulnerability instances together with ground-truth vulnerability metadata, developer-provided fixes, and LLM-generated outputs. Also the CrewAI framework code with agent flows, prompt designs, etc., is provided as an open research material for reproducibility}. Beyond empirical comparison, we discuss the implications of such workflows for secure software engineering, including the continuous need for human-in-the-loop oversight in security-critical settings.

\section{Related Work}

Recent work has shown growing interest in LLM-based multi-agent systems for software engineering more broadly. He et al. \cite{ref_article6} provide a systematic review of 71 primary studies and argue that multi-agent systems can improve autonomy, robustness, and scalability across software engineering tasks, including code generation, quality assurance, maintenance, and end-to-end development workflows. Their review is important in positioning agentic systems as a broader software engineering trend.

A first line of security-oriented work studies multi-agent support for \emph{secure code generation}. AutoSafeCoder \cite{ref_article7} combines a Coding Agent, a Static Analyzer Agent, and a Fuzzing Agent in an iterative loop to improve the security of LLM-generated Python code, reporting a 13\% reduction in vulnerable outputs relative to a baseline model. This work is valuable in showing that role separation and tool-supported feedback can improve security outcomes, but its focus is on secure code generation (only Python). In a related but single-pipeline repair setting, LLM4CVE \cite{ref_article3} proposes an iterative LLM-based vulnerability repair process for real-world CVEs, using repeated feedback, prompt engineering, and Low-Rank Adaptation (LoRA)-enhanced generation to synthesize candidate patches and compare them against trusted fixes using metrics such as CodeBLEU, human quality judgment, and end-to-end compilation. While this work demonstrates the promise of iterative LLM-based repair, it does not explicitly decompose vulnerability handling into distinct roles such as planning, diagnosis, fixing, and verification.

A second line of work focuses on \emph{vulnerability detection}. VulAgent \cite{ref_article1} addresses project-level vulnerability detection through a multi-agent hypothesis-validation pipeline. It uses specialized agents to surface candidate vulnerable code, aggregates findings, constructs explicit vulnerability hypotheses, and validates them against program context, improving average accuracy by 7\% and reducing false positives by about 36\% on PrimeVul and SVEN. This is highly relevant in showing the importance of context-sensitive reasoning and staged analysis for vulnerability detection. However, its scope is detection-centric, it does not extend into patch generation, independent fix verification, or a workflow designed around secure software engineering stages.

A third line of work has emphasized \emph{evaluation settings and workflow realism}. CVE-Bench \cite{ref_article2} argues that existing vulnerability-repair benchmarks are limited because they often reduce the task to simple code input-output pairs. It therefore proposes a more realistic evaluation setup with interactive environments, tool use, and multiple levels of CVE information, including black-box and white-box settings. On 509 CVEs across four languages and 120 repositories, the evaluated agent repairs only about 21\% of cases at best, and performs substantially worse under black-box information settings. In a more workflow-complete direction, MAVM \cite{ref_article9} proposes a multi-agent framework for end-to-end \emph{recurring} vulnerability management, integrating a vulnerability knowledge base with detection, confirmation, repair, and validation, and outperforming baselines by 31.9\%-45.2\% in repair accuracy on recurring vulnerability cases. MAVM is especially relevant because it moves closer to real security workflows, but it targets recurring vulnerability management and patch porting with repository-level historical knowledge rather than file-level vulnerability analysis and mitigation in low-level C code.

In summary, existing work has explored secure code generation, vulnerability detection, iterative repair, realistic benchmarking, and recurring vulnerability management. Several of these studies already demonstrate the value of multi-agent decomposition and tool support. However, there remains limited work on \emph{role-based, workflow-oriented architectures for secure software engineering} that explicitly separate investigation planning, vulnerability diagnosis, patch generation, and fix verification in a way that mirrors industrial security review practice. In addition, prior work has given limited attention to MCP-style tool integration in the analysis stage for low-level C vulnerabilities under a controlled exploratory setup. Our work takes an initial step toward this gap by studying two role-based workflows on CrewAI framework curated on real-world C CVEs, with emphasis on the quality of detections, generated patches, and verification outcomes rather than benchmark-scale throughput.

\section{Methodology: Workflow Design and Evaluation Setup}

We selected some vulnerability instances (CVEs) from the NVD, recorded the associated details, and the vulnerable and patched code files. Later we provided the vulnerable code to our CrewAI based architecture and asked for vulnerability detection and patch generation. Finally, we compared the output with the developer-provided patched version. The details of the workflow design, experimental setup, and analysis procedure are provided in the following sub-sections. In total, the experimental setup yielded 150 workflow executions (25 CVEs × 3 models × 2 workflow variants).

\subsection{Data Collection: CWE, Repository and CVE selection}

We constructed a curated vulnerability set tailored to the needs of this study rather than directly adopting an existing benchmark dataset (CrossVul, BigVul, SVEN, DiverseVul \cite{ref_article10}, etc.). This gave us tighter control over code scope, vulnerable-to-patched version alignment, vulnerability difficulty, and the presence of clearly documented ground-truth. 

We restricted the study to the C programming language in order to keep the evaluation setting controlled and focused on a low-level system language where vulnerability reasoning is often more challenging. Compared to higher-level languages such as Python or Java, C vulnerabilities depend more frequently on memory operations, pointer behavior, bounds handling, and other low-level semantics, making it a suitable setting for studying tool-augmented agentic workflows.

We selected five CWE categories that occur frequently in C-related vulnerability reports: \textbf{CWE-476 (Null Pointer Dereference), CWE-125 (Out-of-bounds Read), CWE-191 (Integer Underflow), CWE-190 (Integer Overflow or Wraparound), and CWE-787 (Out-of-bounds Write).} Selection was guided by the prevalence of these weakness classes in public NVD records \footnote{\url{https://nvd.nist.gov/general/visualizations/vulnerability-visualizations/cwe-over-time}} and MITRE 2025 CWE Top 25 list \footnote{https://cwe.mitre.org/top25/archive/2025/2025\_cwe\_top25.html} and by their relevance to common low-level security issues in C programs. We then identified repositories that were widely used and had a significant number of reported vulnerabilities. Table \ref{tab:repo_summary} summarizes the repositories considered, together with their associated CVEs counts as per NVD, GitHub popularity indicators, and the number of CVEs considered from each repository for our study.

\begin{table}[t]
\caption{Repositories considered for CVE selection.}
\label{tab:repo_summary}
\centering
\scriptsize
\begin{tabular}{p{4.5cm}p{1.8cm}p{1.8cm}p{2.5cm}}
\hline
\textbf{Repository} & \textbf{NVD CVEs} & \textbf{GitHub stars} & \textbf{CVEs Selected for study} \\
\hline
freerdp/freerdp & 167 & 13.1k & 5 \\
nanomq/nanomq & 33 & 2.5k & 4 \\
AcademySoftwareFoundation/\\openexr & 84 & 1.8k & 8 \\
wolfSSL/wolfssl & 107 & 2.8k & 8 \\
\hline

\end{tabular}
\end{table}

Within these repositories, we searched for CVEs belonging to the selected CWE categories and applied the following inclusion criteria:
\begin{enumerate}
    \item We prioritized \emph{recent vulnerability} instances in order to keep the study aligned with contemporary coding practices and vulnerability reporting.
    
    \item We retained only those CVEs for which both the \emph{vulnerable code} version and the corresponding \emph{patched version} could be reliably recovered.
    
    \item We selected cases with sufficiently \emph{clear public documentation of the vulnerability}, including its nature, likely root cause, exploitation path, and security impact, so that workflow outputs could be meaningfully compared against available ground truth.
    
    \item To avoid restricting the evaluation to small or narrowly scoped examples, we selected CVEs whose affected files exhibited substantial \emph{variation in size}, ranging from approximately 700 to 20,000 lines of code. We also ensured \emph{diversity in vulnerability localization}, including both cases where the weakness was concentrated in a single line or operation and cases where it extended across multiple lines of code. This allowed the study to examine workflow behavior under different levels of contextual and structural complexity.
    
    \item Finally, we limited the number of instances to five CVEs per CWE category, yielding a total of 25 CVEs, because the goal of this study is not large-scale statistical benchmarking but an initial exploratory assessment of the quality of workflow outputs across representative vulnerability cases.

\end{enumerate}

The complete list of 25 CVEs, together with their metadata (including CVE ID, CWE ID, vulnerable file name, vulnerable line numbers, severity, security impact, and exploitation path when available), as well as the corresponding vulnerable and patched code files, is provided in the replication repository\footnote{https://github.com/SriAbir/SecurityAgentWorkflow}. This study is intentionally conducted at the vulnerable file level rather than the full repository level. At the next level we would extend this to a repository level to understand how does the presence of the entire repository information and structure affect the end results (keeping all other settings intact).


\subsection{Model Selection}

We prioritized open-source LLMs in this study. This choice was motivated not only by reproducibility and deployability considerations, but also by prior discussions with cybersecurity practitioners in industrial settings, which indicated reservations about sending potentially sensitive source code to proprietary hosted models. Locally deployable or openly controllable models were viewed as more compatible with internal security and compliance expectations. We initially began with a pool of 25 candidate models, selected from publicly discussed coding and reasoning capable LLMs \footnote{https://ollama.com/search?q=llama}. Next, we used Inspect Evals’ built-in CyberMetric task (\texttt{cybermetric\_80}) \footnote{https://github.com/CyberMetric} as a preliminary feasibility filter to identify models with sufficient cybersecurity-oriented knowledge for downstream use in the workflow. Based on this step, we retained the top 8 models as presented in Table \ref{tab:model_screening}. From this set, we selected three final LLMs, \texttt{nemotron-cascade-2:30b}, \texttt{qwen3-coder-next}, and \texttt{gpt-oss:120b} while ensuring that all selected models supported MCP-based tool integration. This enabled consistent comparison across workflow variants. To keep it simple, during each execution we used the same model to instantiate all the three roles (Vulnerability Analyzer, Fixer and verifier).

\begin{table}[t]
\caption{Preliminary model screening}
\label{tab:model_screening}
\centering
\scriptsize
\begin{tabular}{p{4.0cm}p{1.5cm}p{1.5cm}p{2.4cm}}
\hline
\textbf{Model} & \textbf{Accuracy} & \textbf{Stderr} & \textbf{Skill Integration} \\
\hline
gpt-oss:120b            & 0.988 & 0.012 & Yes \\
qwen3-coder-next:latest & 0.950 & 0.025 & Yes \\
nemotron-cascade-2:30b  & 0.975 & 0.018 & Yes \\
llama3:70B              & 0.950 & 0.025 & No  \\
gemma4:31b              & 0.988 & 0.012 & No  \\
deepseek-r1:32b         & 0.988 & 0.012 & No  \\
glm-4.7-flash:bf16      & 0.950 & 0.025 & No  \\
qwen3.6:35b             & 0.975 & 0.018 & Yes \\

\hline
\end{tabular}
\end{table}

\subsection{Role-Based Workflow}

\begin{figure}[ht]
\begin{center}
\includegraphics[scale=.30]{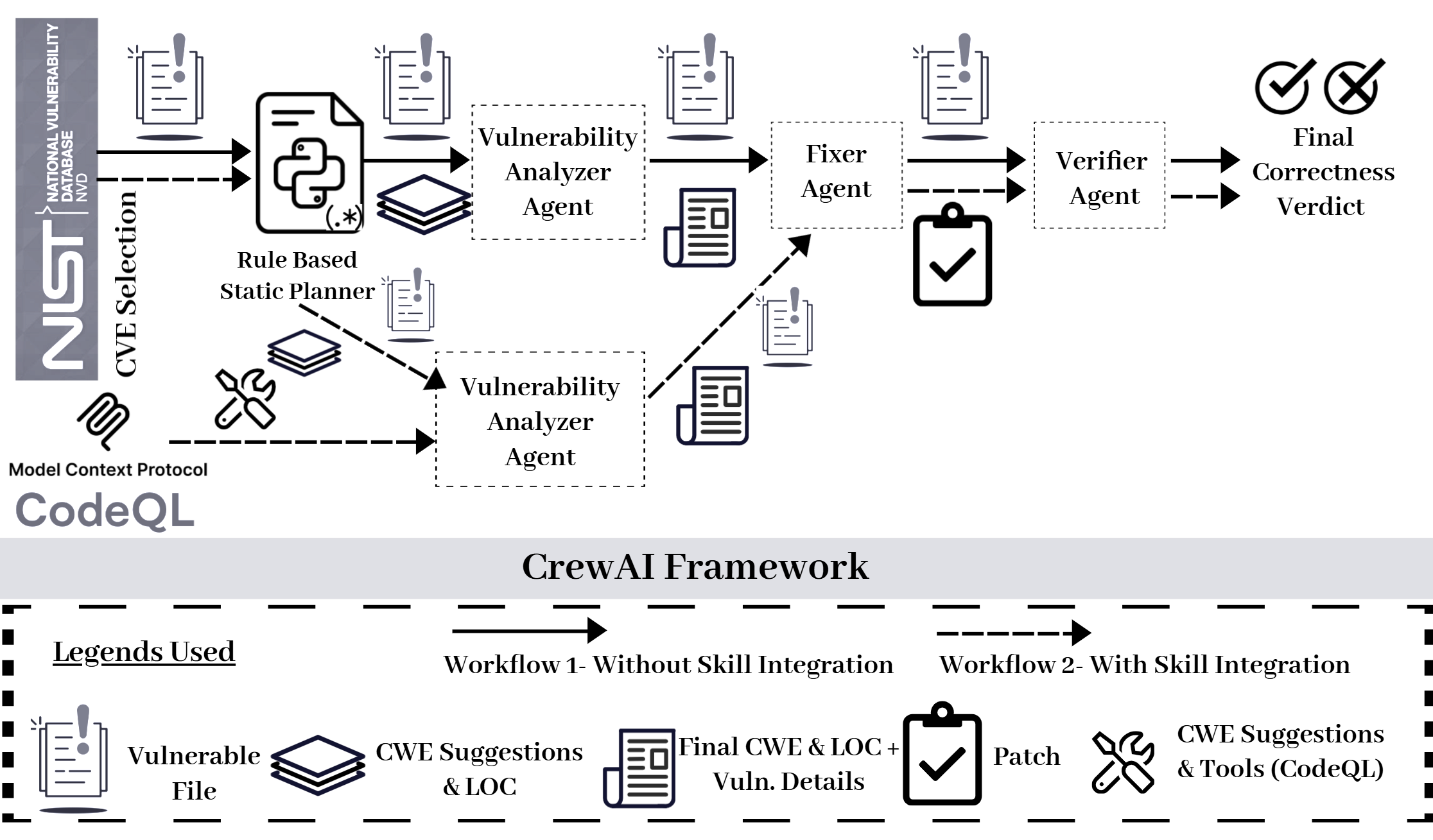}
\caption{Role based Workflow for Secure Software Development}
\label{workflow}
\end{center}
\end{figure}

We implement the proposed workflow in CrewAI \footnote{https://docs.crewai.com/} as a base layer because its explicit role-based orchestration model and easy skill/tool integration settings aligns well with our security workflow design and additional MCP skill customization. In this design, CrewAI is used to coordinate role assignment, task sequencing, and context passing across workflow stages. Our contribution is not a modification of CrewAI itself, but the design and exploratory study of a security-oriented workflow instantiated on top of it. The workflow as depicted in Fig. \ref{workflow} is organized around four roles, \textit{Planner, Vulnerability Analyzer, Fixer, and Verifier}, each with a narrowly scoped responsibility and a restricted output format. The workflow has two variants, the first one (Workflow 1) without skill integration (Planner, Analyzer, Fixer, Verifier) is described as follows.

\begin{figure}[ht]
\begin{center}
\includegraphics[scale=.20]{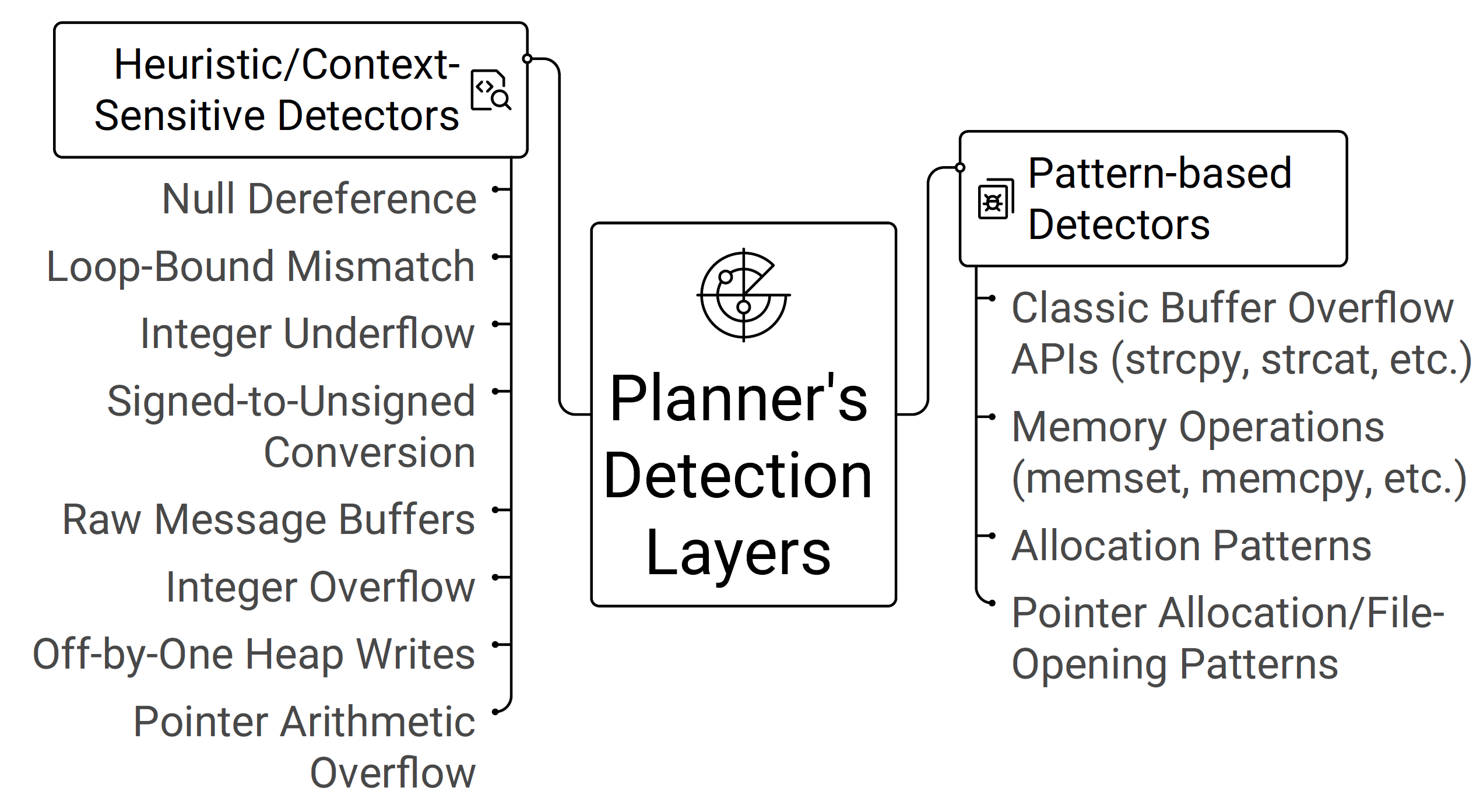}
\caption{Planner Detection Logic}
\label{planner}
\end{center}
\end{figure}

\begin{itemize}
    \item The Planner serves as the investigation-guidance stage. 
    In the present study, this role is implemented as a rule-based static planner rather than an LLM agent. The planner (.py file) scans the input C code for predefined vulnerability-relevant patterns (Fig. \ref{planner}) and highlights likely investigation regions together with corresponding analysis focus. It uses regular-expression-based matching to identify patterns associated with common C weakness classes, including classic buffer overflow operations, out-of-bounds memory access risks, potential null pointer de-references, and integer-overflow/underflow related allocation issues. It returns detected issue regions and a set of prioritized vulnerability themes, which are then used only as guidance for the downstream Analyzer. 

    \item The Vulnerability Analyzer is an LLM based agent that performs the main technical diagnosis of the vulnerable code. It receives the source code together with the planning context and is instructed to determine the primary vulnerability, assign the CWE, identify the vulnerable lines of code, explain the root cause, trace the source-to-sink or input-to-dangerous-operation flow, and characterize the security impact. To preserve role separation, the Analyzer is explicitly constrained to use the Planner output only as guidance and not to restate the plan or propose a patch. This makes the Analyzer the principal decision-making stage of the workflow.

    \item The Fixer (LLM-based) is responsible for remediation. Given the vulnerable source code and prior analysis context, it produces a minimal secure patch targeted at the vulnerability diagnosed by the Analyzer. Its output is restricted to the patched code together with a concise explanation of the unsafe behavior removed, the security control or validation added, and why the patch addresses the root cause. The Fixer is deliberately prevented from redoing the entire diagnosis, which allows the study to separate vulnerability understanding from patch construction.

    \item The Verifier agent acts as an independent post-remediation assessment stage. Given the original source code and prior task context, it evaluates whether the proposed patch actually resolves the vulnerability identified by the Analyzer. The Verifier is instructed to assess whether the same vulnerability was addressed, whether the root cause rather than only the symptom was removed, whether the unsafe source-to-sink path was disrupted, whether attacker-controlled input can still reach the dangerous operation, and whether residual risk or correctness regressions remain. It returns a structured verification verdict together with patch-level concerns.
    
\end{itemize}

\vspace{-3em}
In the second variant i.e., Workflow 2, (Planner, Analyzer with MCP CodeQL integration, Fixer, Verifier), MCP CodeQL \footnote{https://mcpmarket.com/server/codeql} integration equips the Analyzer with an additional static-analysis skill during the diagnosis stage. CodeQL is particularly appropriate in our setting because it provides built-in support for C and C++, including dedicated C/C++ library, standard query packs, and security and quality query suites for static analysis \cite{ref_article11}. We relied on CodeQL's built-in C/C++ security analysis capabilities exposed through the MCP integration and did not develop custom CodeQL queries for this study. The Analyzer records the CodeQL MCP tools it uses and the contribution of each tool to the diagnosis, but remains responsible for validating or rejecting that evidence against the actual code. 

We present the role descriptions and workflow logic at a conceptual level in this paper. The full role prompts, task specifications, and implementation details are provided in the accompanying replication repository \footnote{https://github.com/SriAbir/SecurityAgentWorkflow}.

\subsection{Execution Environment and Run Settings}

The experiments were executed on two servers, 1) Server 1 with Windows was equipped with an AMD Ryzen 9 9950X CPU, 128 GB RAM, and an NVIDIA RTX 5090 GPU, 2) Server 2 ran on Linux and used an NVIDIA GB10 platform with 128 GB RAM and 128 GB unified memory along with GB 10 NVIDIA GPU. The workflow implementation was developed in Python 3.10.12 using CrewAI v1.14.2.

To keep the experimental conditions consistent, the same role prompts were used across all selected models within each workflow variant. Likewise, the same static Planner was used in both workflow variants, so that the only architectural difference between them was the Analyzer’s access to MCP CodeQL in Workflow 2. This allowed the comparison to focus on the effect of tool-augmented analysis rather than changes in workflow structure. Each CVE instance was executed twice for each model and workflow variant. Across these repeated runs, we did not observe major differences in the outcomes. In particular, cases that failed to give the correct solution in the first run did not become succeed in the second run, and cases that succeeded remained stable. For this reason, we do not report per-run details separately and instead discuss the consolidated outcomes. The temperature was set to 0.5 for all model runs. No manual editing was performed on intermediate agent outputs during workflow execution. Outputs generated by the Planner, Analyzer, Fixer, and Verifier were passed between stages according to the predefined task structure. 

\subsection{Data Analysis and Evaluation}

\subsubsection{Ground Truth Definition}
For each selected CVE, ground truth was derived from the publicly available vulnerability description, primarily by following the advisory URL referenced on the corresponding NVD page, together with the associated repository artifacts. For the \emph{Vulnerability Analyzer}, the reference information consisted of: i) the associated CWE type, ii) the vulnerability location (i.e., the lines containing the vulnerable logic), iii) the root-cause description, iv) the exploitation path when available, and v) the security impact. For the \emph{Fixer}, the primary ground truth was the developer-provided patched code, evaluated semantically rather than by exact textual matching. For the \emph{Verifier}, ground truth was defined by whether the generated patch correctly removed the vulnerability identified in the reference case, broke the unsafe path or condition that enabled it, and avoided introducing obvious residual risk or regression.

\begin{figure}[ht]
    \centering

    \begin{subfigure}{\linewidth}
    \centering
    \includegraphics[scale=0.30]{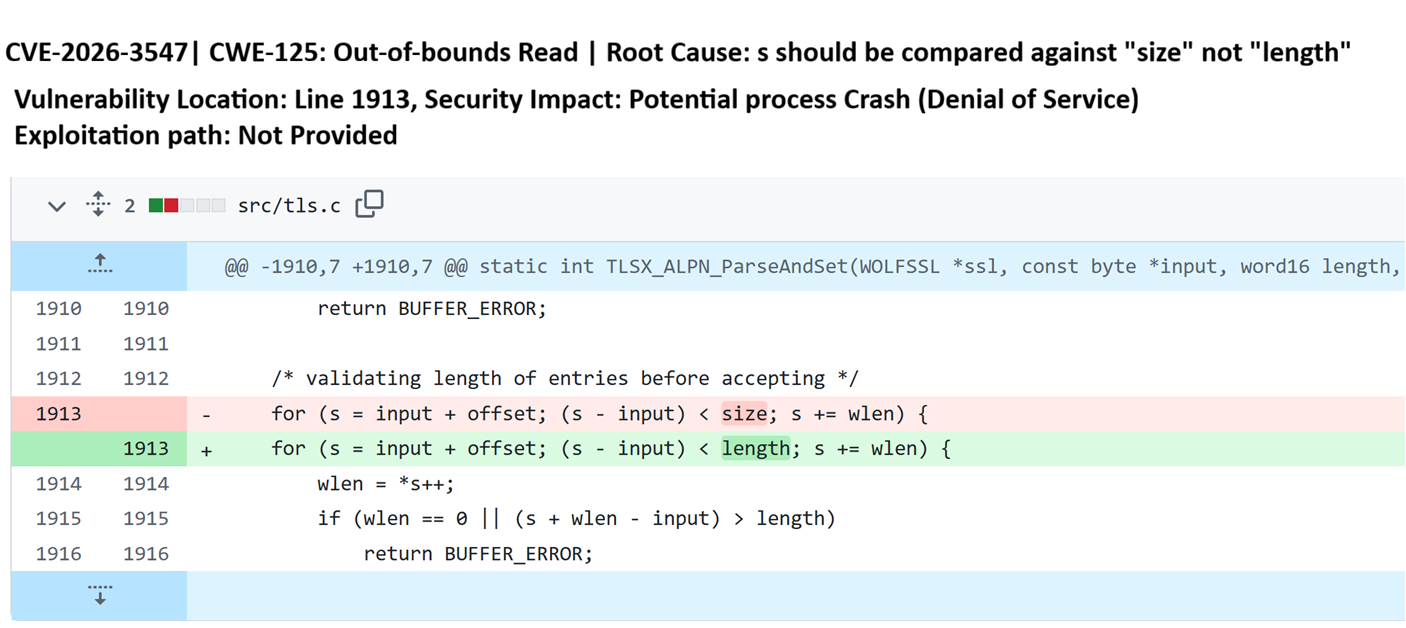}
    \caption{Ground-truth for vulnerability \& patch for CVE-2026-3547.}
    \label{fig:cve3547-groundtruth}
\end{subfigure}
    \begin{subfigure}{\linewidth}
        \centering
        \includegraphics[scale=0.30]{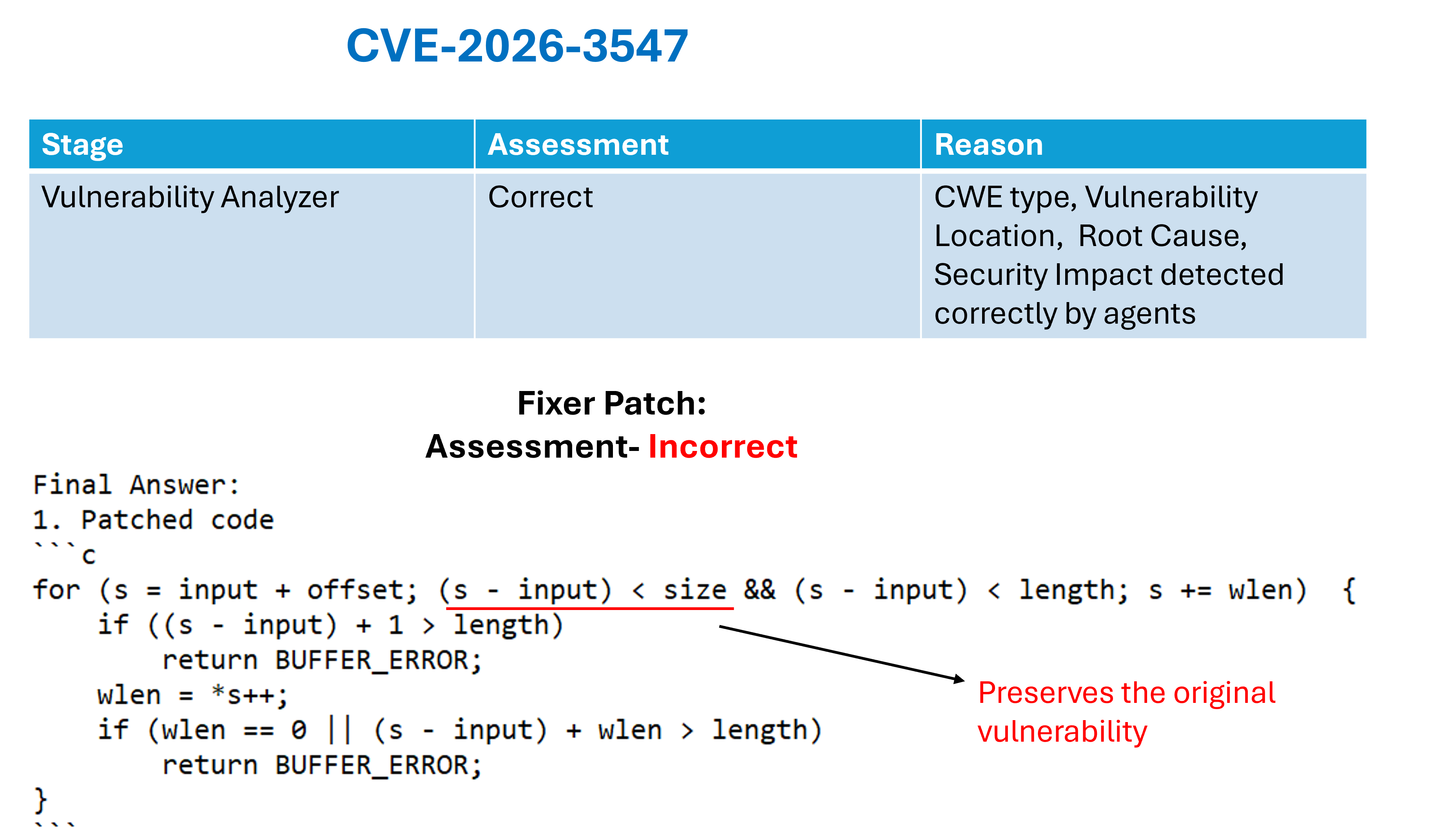}
        \caption{Assessment of an agent-generated patch. Although the agent correctly identified the vulnerability, its fix retained \texttt{size} in the loop condition, thereby preserving the original boundary-checking flaw.}
        \label{fig:cve3547-assessment}
    \end{subfigure}
    \caption{Example: Ground Truth and Assessment for CVE-2026-3547}
    \label{fig:cve3547-evaluation-example}
\end{figure}

\subsubsection{Rubric-Based Manual Assessment}
The outputs produced by the workflow were evaluated through a \emph{manual, rubric-based assessment} against the reference ground truth defined above. Two authors independently did the assessment, and inter-rater agreement of $\kappa$ = 0.83 (almost perfect agreement) was measured using Cohen's kappa \cite{cohen}. All disagreements were then resolved through discussion, and the agreed labels were used in the final analysis. This approach was adopted because the study focuses on the semantic quality of security reasoning and patching outcomes rather than on exact textual similarity. In particular, multiple secure outputs may differ syntactically while still addressing the same underlying vulnerability. Accordingly, each workflow stage was assessed using predefined criteria and mapped to a small set of qualitative labels.

For the \emph{Vulnerability Analyzer}, we assessed the correctness of, i) CWE assignment, ii) vulnerable line(s) or code region, iii) quality of the root-cause explanation, iv) reported exploitation or input-to-dangerous-operation flow, and v) security impact. Based on these criteria, Analyzer outputs were labeled as \emph{Correct}, \emph{Partially Correct}, or \emph{Incorrect}. For the \emph{Fixer}, we assessed: i) whether the generated patch addressed the vulnerability diagnosed in the reference case, ii) whether the patch was semantically comparable to the trusted developer fix, and iii) whether it introduced obvious unsafe behavior, regressions, or overly broad modifications. Based on these criteria, Fixer outputs were labeled as \emph{Comparable}, \emph{Partially Correct}, \emph{Incorrect}, and \emph{Better}. For the \emph{Verifier}, we assessed: i) whether the reported verdict (\emph{fixed}, \emph{partially fixed}, or \emph{not fixed}) was consistent with the reference case, ii) whether the reasoning correctly judged root-cause removal, and iv) whether residual risk or possible regressions were meaningfully identified. Based on these criteria, Verifier outputs were labeled as \emph{Accurate}, and \emph{Inaccurate}.

Table \ref{tab:rubric} presents the rubrics for labeling the workflow outputs. Additionally, Fig. \ref{fig:cve3547-evaluation-example} presents an example of ground truth and  assessment.
\begin{table}[t]
\caption{Rubric used for manual assessment of workflow outputs.}
\label{tab:rubric}
\centering
\scriptsize
\begin{tabular}{p{2.1cm}p{4.8cm}p{5.1cm}}
\hline
\textbf{Stage} & \textbf{Assessment} & \textbf{Decision rule} \\
\hline

\multirow{3}{*}{Analyzer}
& Correct
& When all the 5 assessment criteria matches with ground truth \\
\cline{2-3}
& Partially Correct
& The vulnerable line(s) of code are identified correctly but one or more of the other criteria do not match ground truth OR CWE-ID and root cause identified correctly but one or more of the other criteria do not match ground truth\\
\cline{2-3}
& Incorrect
& Both the CWE-ID and vulnerable line(s) of code are identified incorrectly\\
\hline

\multirow{3}{*}{Fixer}
& Comparable
& The patch is semantically similar to ground truth \\
\cline{2-3}
& Partially Correct
& The patch contains semantically similar logic to the ground truth but introduces additional buggy logic, inappropriate for the code \\
\cline{2-3}
& Incorrect
& The patch is totally different from ground truth and does not resolves the vulnerability at all\\
\cline{2-3}
& Better
& The patch contains additional security controls over the ground truth which is suitable for improving the code security \\
\hline

\multirow{3}{*}{Verifier}
& Accurate
& The verifier gives the verdict which is similar to our evaluation after comparison with ground truth \\
\cline{2-3}
& Inaccurate
& The verifier gives the verdict which is different from our evaluation after comparison with ground truth \\
\hline

\end{tabular}
\end{table}

\section{Results \& Implications}


We report overall Analyzer, Fixer, and Verifier accuracy as the percentage of cases in which each role produced the correct outcome relative to the ground truth. While significant observations are highlighted here, the detailed results are present in the shared repository.

\paragraph{\textbf{Vulnerability Analyzer:}}

\begin{figure}[ht]
\begin{center}
\includegraphics[scale=.3]{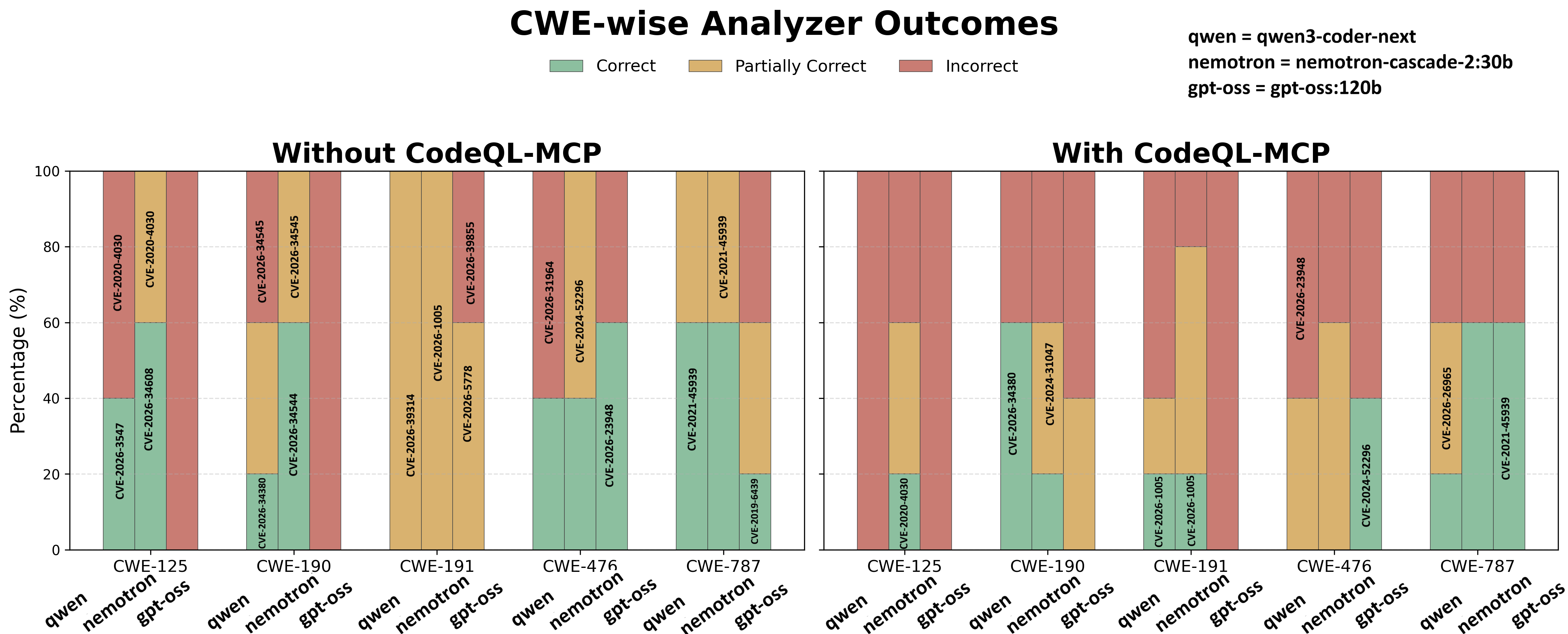}
\caption{Results: Vulnerability Analyzer}
\label{fig:analyzer}
\end{center}
\end{figure}

Figure \ref{fig:analyzer} shows a clear dependence of analyzer performance on both the model and the availability of CodeQL-MCP support, with notable variation across CWEs.

Without CodeQL-MCP, \emph{nemotron-cascade-2:30b} is the strongest overall analyzer. It achieves the highest proportion of fully correct vulnerability detections for \emph{CWE-125, CWE-190, and CWE-787}. \emph{qwen3-coder-next} performs moderately well on \emph{CWE-476 and CWE-787}, but is weaker on \emph{CWE-191}, where its outputs are mostly only partially correct. \emph{gpt-oss:120b} is the weakest analyzer in the no-CodeQL setting.

With CodeQL-MCP, the impact is mixed rather than uniformly positive. \emph{qwen3-coder-next} improves on \emph{CWE-190} and shows some gains on \emph{CWE-191}, but remains weak on \emph{CWE-125 and CWE-476}. \emph{nemotron-cascade-2:30b} remains the most balanced model overall, with strong results on \emph{CWE-190 and CWE-787}, and a reasonable mix of correct and partially correct outcomes on \emph{CWE-191 and CWE-476}. In contrast, gpt-oss:120b continues to struggle even with CodeQL-MCP, with a small improvement for \emph{CWE-787}.

Across CWE categories, \emph{CWE-787 and CWE-476} appears to be the easiest class overall, producing the largest share of correct localizations across models and settings. While \emph{CWE-191} is the hardest, with very few fully correct results. \emph{CWE-125} also remains challenging, especially for \emph{qwen3-coder-next and gpt-oss:120b} in the CodeQL-assisted setting. Overall, the results suggest that \emph{CodeQL-MCP} helps selectively rather than universally, and that \emph{nemotron-cascade-2:30b} gives the best accuracy across weakness types under our controlled experimental settings.

The analyzer results suggest that most failures arose from two main sources: (i) the Planner failed to surface the required vulnerability pattern, accounting for \textbf{16\%} of cases, and (ii) even when the Planner provided the correct signal, the Analyzer often struggled to prioritize the most relevant vulnerability cue. A small number of failures were also linked to misleading CodeQL outputs that diverted the analysis. Exact vulnerable-region detection reached \textbf{52\%} accuracy in without CodeQL-MCP workflow and \textbf{56\%} in with CodeQL-MCP workflow. In a few instances, limited code context may also have contributed to failure, suggesting that full-repository access could further improve analyzer performance.

\paragraph{\textbf{Fixer:}}

\begin{figure}[ht]
\begin{center}
\includegraphics[scale=.30]{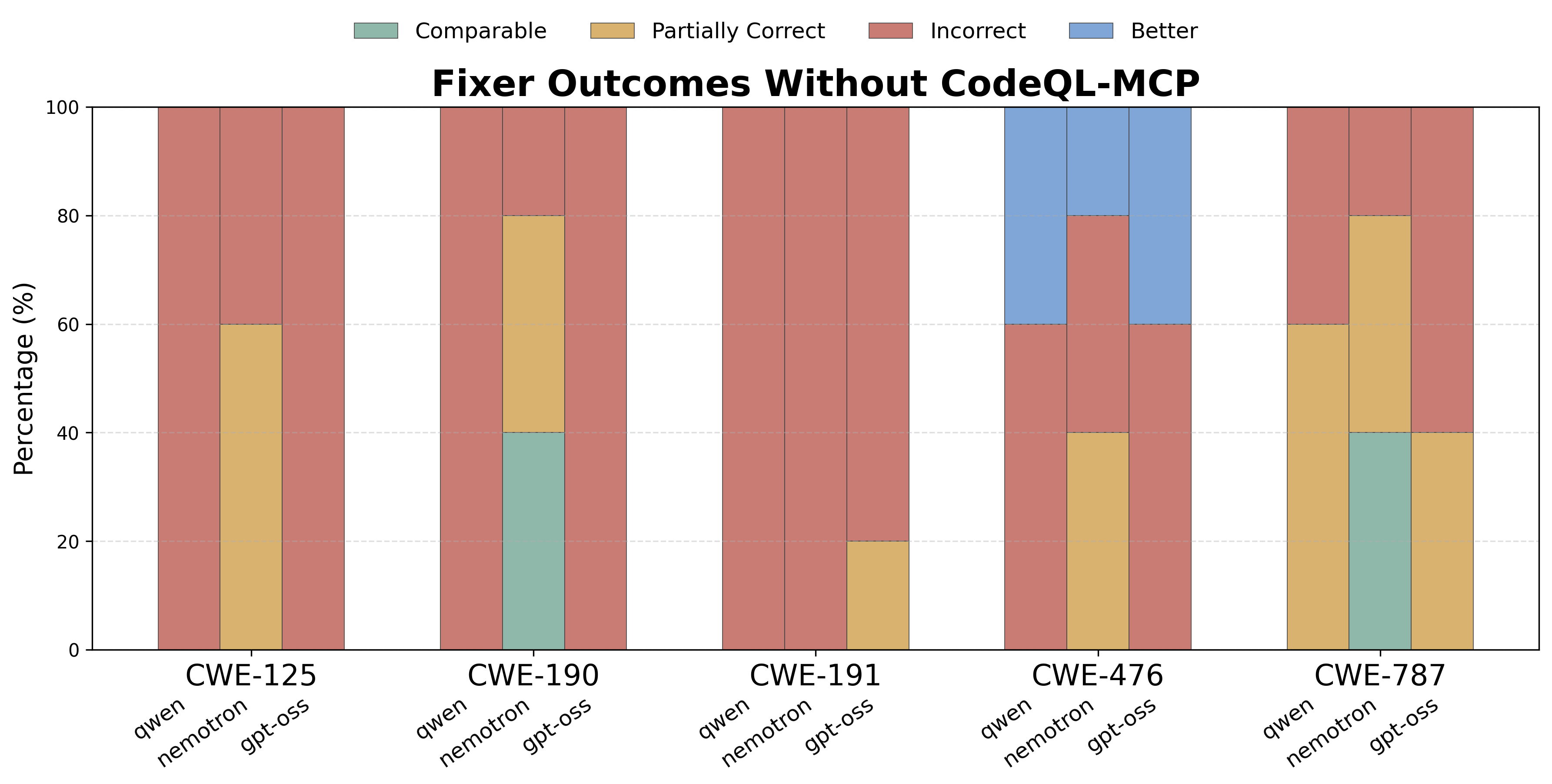}
\caption{Results: Fixer}
\label{fig:fixer}
\end{center}
\end{figure}

The fixer results (Figure \ref{fig:fixer}) indicate that repair remains substantially more difficult than detection. 

Among the three models, \emph{nemotron-cascade-2:30b} shows the most balanced fixer behavior, producing some comparable or partially correct repairs across multiple CWEs. The strongest outcomes appear for \emph{CWE-476}, where some fixes are actually better than the ground truth, and for \emph{CWE-787}, where \emph{nemotron-cascade-2:30b} achieves the highest proportion of comparable repairs. In contrast, \emph{CWE-191} remains the most difficult category.

\vspace{-3em}

Since fixer accuracy did not vary much between the with-CodeQL and without-CodeQL settings (expected, since CodeQL was integrated with the analyzer rather than the fixer) we presented the without-CodeQL results in Figure \ref{fig:fixer}. 

In case of the Fixer agent, \textbf{24\%} of the generated patch contained additional logic in addition to what was actually required to mitigate the vulnerability. Additionally, under such cases, we observed instances where the additional logic was of no harm (though not required) as well as certain instances where these unnecessary updates resulted in to buggy logic.

\paragraph{\textbf{Verifier:}}
Verifier accuracy was highest for \textit{nemotron-cascade-2:30b} (\textbf{69\%}), followed by \textit{qwen3-coder-next} (\textbf{44\%}) and \textit{gpt-oss:120b} (\textbf{38\%}).
\vspace{-1em}

\paragraph{\textbf{Baseline, Ablation Studies \& Summary:}}

Table~\ref{tab:final_accuracy_summary} summarizes the overall detection and fix accuracy across the evaluated workflows (all of them with \textit{nemotron-cascade-2:30b}) and the baseline setting. We measure accuracy as the percentage of cases that match the ground truth, with separate scores for vulnerability detection and fix generation. Within our evaluated models and settings, Workflow 1 with \textit{nemotron-cascade-2:30b} performs best (comparable to GPT 5.5 in detection accuracy), the Planner is critical for detection quality, and CodeQL support has a larger effect on analysis than on fixing.


\noindent \textbf{Answer- RQ1 \& RQ2}: The role-based agentic workflow  without CodeQL using \textit{nemotron-cascade-2:30b}, achieved \textbf{44\%} detection accuracy which was comparable to GPT 5.5, while fix accuracy was lower at \textbf{19\%} and verifier reached \textbf{69\%} accuracy. Removing the Planner reduced correct detections by \textbf{45\%}, underscoring the importance of explicit planning and role separation. Workflow~2 with CodeQL yielded \textbf{31\%} detection accuracy and \textbf{13\%} fix accuracy. These findings indicate that tool integration alone is insufficient; its benefit depends on whether the Analyzer can correctly interpret and prioritize the additional hints.

\vspace{-2em}
\begin{table}[ht]
\centering
\scriptsize
\caption{Final accuracy summary across workflows with Baseline Comparison}
\label{tab:final_accuracy_summary}
\begin{tabular}{lcc}
\toprule
\textbf{Workflow / Setting} & \textbf{Detection Accuracy (\%)} & \textbf{Fix Accuracy (\%)} \\
\hline
Workflow 1 (without CodeQL) & 44 & 19 \\
\hline
Workflow 2 (with CodeQL) & 31 & 13 \\
\hline
Workflow 1 (without CodeQL) (without Planner)              & 25 & 13 \\
\hline
Workflow 2 (with CodeQL) (without Planner)               & 19 & 6 \\
Baseline (\textit{OpenAI GPT-5.5.})                   & 44 & 44 \\

\bottomrule
\end{tabular}
\end{table}
\vspace{-1em}

\section{Discussion}

Our discussion is mainly based on practitioner insights with whom we had a detailed session where we presented these results. We wanted to understand how our approach would be beneficial to them and in what ways they would like to apply or integrate our framework in their secure software engineering process. The highlights of the discussions are as follows:

\begin{itemize}
    \item It was interesting to see how the planners could directly influence the Analyzer's approach. It could be better to have a human intervention at this step (right after planner) who could prioritize the planner hints before we send the same to the Analyzer
    \item Though the Fixer gave some partially correct patch, it could still be beneficial for the practitioners. E.g., if 3 out of 5 statements added by the Fixer are correct, a human-in-the-loop process could be followed to filter out the correct remediation steps and ignore the others.
    
    \item It is a challenge to cope up with the everyday releases of newer and more powerful models. It is hard to understand which could be better over the other. Therefore, through our work we try to shift the focus on the underlying role-based workflow rather than the models. Though an initial screening for filtering out some candidate models is essential but the main aim s to refine and finetune the architecture and integration techniques.

  \item Finally, though the current CodeQL skill integration did not heavily improve the performance of the system, there still remains ample scope to modify the way the MCP skill results are being used or explore other skills in the workflow.     
\end{itemize}

\section{Threats to Validity}
\vspace{-1em}
\label{sec:validity}
Following the validity perspectives discussed by Wohlin \cite{ref_18} and Staron \cite{ref_19}, we summarize the main threats as follows.

\textbf{Construct validity.} The study evaluates workflow quality through a manual, rubric-based assessment of analyzer, fixer, and verifier outputs. Although this is suitable for judging semantic adequacy in a small exploratory study, some subjectivity remains in assigning qualitative labels such as \emph{Correct}, \emph{Partially Correct}, or \emph{Comparable}. To reduce this threat, we defined explicit evaluation criteria and grounded the assessment in publicly available CVE descriptions, associated CWE labels, vulnerable code regions, and trusted patched versions.

\textbf{Internal validity.} The workflow behavior may be influenced by model stochasticity and tool interaction effects. To reduce this threat, we used the same prompts across models within each workflow, kept the same static Planner in both workflow variants, and repeated each CVE execution twice. Since repeated runs did not materially alter the outcome categories, results are reported at the consolidated case level.

\textbf{External validity.} The study uses a curated set of 25 real-world C-language CVEs across five CWE categories. While this supports controlled exploratory evaluation, it limits generalizability to other vulnerability classes, programming languages, and repository-scale workflows. The findings should therefore be interpreted as early evidence rather than broad statistical claims.

\textbf{Conclusion validity.} The study relies on a limited set of three LLMs and one tool-augmented variant using MCP CodeQL integration. Although the workflow design, role prompts, and artifacts are provided in the replication repository, broader replication across additional models and tool configurations is needed to assess stability fully.

\section{Conclusion and Future Work}
\vspace{-1em}

This paper explored a role-based agentic workflow for vulnerability analysis and mitigation in C/C++. The results suggest that explicit role separation with a strongly designed planner can support improved vulnerability detection. Also, CodeQL-based tool support showed mixed effects denoting that tool-integration needs to be done in a way that can be prioritized and interpreted by the agents as per requirement. At the same time, the uneven quality of generated patches reinforces the continued need for human-in-the-loop review in security-critical settings. 

These findings position workflow-oriented agentic architectures as a promising direction for security-focused software process improvement, while motivating future studies on larger datasets and broader tool and model settings. Future studies is specifically aimed towards designing different workflow alternatives (with multiple analyzers, judges, refinement loops, etc.). Also a repository level analysis in future could help improve the detection and mitigation process (more code context and transparent inter-file relations).

\begin{credits}
\subsubsection{\ackname}  This paper has been partially financed by Software Center,
www. software-center.se, a collaboration between Chalmers, the University of Gothenburg, and 17 companies. The paper has also been partially funded by the SFO Transport/AoA Transport at the University of Gothenburg and Chalmers University of Technology.

\end{credits}
%
%
%
%

\end{document}